\documentclass[12pt]{article}

\usepackage{amsmath}
\DeclareMathAlphabet\mathbb  {U}{msb}{m}{n}
\DeclareFontFamily{U}{msb}{} \DeclareFontShape{U}{msb}{m}{n}{
  <5> <6> <7> <8> <9> gen * msbm
  <10> <10.95> <12> <14.4> <17.28> <20.74> <24.88> msbm10
  }{}

\renewcommand{\title}[1]{\null\vspace{10mm}\noindent
                         {\Large{\bf #1}}\vspace{10mm}}
\newcommand{\authors}[1]{\noindent{\large #1}\vspace{5mm}}
\newcommand{\address}[1]{\center{\noindent\small\itshape #1\vspace{0mm}}}

\makeatletter

\def\@citex[#1]#2{\if@filesw\immediate\write\@auxout
        {\string\citation{#2}}\fi
\def\@citea{}\@cite{\@for\@citeb:=#2\do
        {\@citea\def\@citea{,}\@ifundefined
        {b@\@citeb}{{\bf ?}\@warning
        {Citation `\@citeb' on page \thepage \space undefined}}
        {\csname b@\@citeb\endcsname}}}{#1}}
\newif\if@cghi
\def\cite{\@cghitrue\@ifnextchar [{\@tempswatrue
        \@citex}{\@tempswafalse\@citex[]}}
\def\citelow{\@cghifalse\@ifnextchar [{\@tempswatrue
        \@citex}{\@tempswafalse\@citex[]}}
\def\@cite#1#2{{\if@cghi\unskip$\null^{#1}$\else
#1\fi\if@tempswa\typeout
        {warning: optional citation argument ignored: `#2'} \fi}}

\def\section{\@startsection{section}{1}{\z@}{-3.25ex plus -1ex minus
             -.2ex}{1.5ex plus .2ex}{\normalfont\bfseries}}
\def\subsection{\@startsection{subsection}{1}{\z@}{-3.25ex plus -1ex
                minus -.2ex}{1.5ex plus .2ex}{\normalfont\itshape}}

\renewenvironment{thebibliography}[1]
         {\section*{References}\frenchspacing\small
          \begin{list}{\arabic{enumi}.}
         {\usecounter{enumi}\parsep=2pt\topsep 0pt
         \settowidth{\labelwidth}{[#1]}
         \leftmargin=\labelwidth\advance\leftmargin\labelsep
         \rightmargin=0pt\itemsep=0pt\sloppy}}{\end{list}}

\makeatother

\def\smallhatbarlambda{\raisebox{-1mm}{\scriptsize$\hat{\bar{%
                  \raisebox{1mm}{\scriptsize$\lambda$}}}$}}

\begin{document}

\allowdisplaybreaks[2]
\thispagestyle{empty}

\begin{titlepage}

\begin{center}
\hspace*{\fill}{{\normalsize \begin{tabular}{l}
                              \textsf{hep-th/0205094}
                              \end{tabular}   }}

\title{Seiberg-Witten map for noncommutative \\[1mm]
super Yang-Mills theory}

\authors {{V.~Putz$^{*}$, R.~Wulkenhaar$^{\dag}$}}

\address{Max-Planck-Institute for Mathematics in the Sciences\\
Inselstra\ss{}e 22--26, D-04103 Leipzig, Germany}      

{\renewcommand{\thefootnote}{\fnsymbol{footnote}}

\footnotetext[1]{vputz@mis.mpg.de, work supported by ``Fonds zur F\"orderung
der Wissenschaftlichen Forschung'' (FWF) under contract P15015-TPH.}

\footnotetext[2]{rwulken@mis.mpg.de, Schloe\ss{}mann fellow}

}
\vspace{10mm}

\begin{minipage}{12cm}
  
  {\it Abstract.} In this letter we derive the Seiberg-Witten map for
  noncommutative super Yang-Mills theory in Wess-Zumino gauge.
  Following (and using results of) \texttt{hep-th/0108045} we split
  the observer Lorentz transformations into a covariant particle
  Lorentz transformation and a remainder which gives directly the
  Seiberg-Witten differential equations. These differential equations
  lead to a $\theta$-expansion of the  noncommutative super Yang-Mills
  action which is invariant under commutative gauge transformations
  and commutative observer Lorentz transformation, but not invariant
  under commutative supersymmetry transformations: The
  $\theta$-expansion of noncommutative supersymmetry leads to a
  $\theta$-dependent symmetry transformation. For this reason the
  Seiberg-Witten map of super Yang-Mills theory cannot be expressed in
  terms of superfields. 
  
\vspace*{1cm}
\end{minipage}

\end{center}

\end{titlepage}
\renewcommand{\thefootnote}{\alph{footnote}}              

\section{Introduction}

The simplest model for noncommutative space-time is the so-called
noncommutative $\mathbb{R}^4$ characterised by a $\star$-product
involving a constant antisymmetric tensor $\theta$. Field theories on
such a deformed space-time became recently very popular, mainly due to
their relation to string theory \cite{Seiberg:1999vs} and the
possibility to perform similar calculations of Feynman graphs as on
usual commutative space-time. It turned out that field theories which
are renormalisable in the commutative world are in general not
renormalisable\cite{Chepelev:2001hm} (at any loop order) on
noncommutative $\mathbb{R}^4$.

One may ask then whether expanding the $\star$-product in $\theta$
improves the renormalisability\footnote{In spite of encouraging
  preliminary results \cite{Bichl:2001cq}, it was shown that
  $\theta$-expansion does not help \cite{Wulkenhaar:2001sq}.}. If the
action of the field theory has a symmetry on noncommutative level, the
$\theta$-expanded symmetry transformation will, in general, mix the
orders of the $\theta$-expanded action. A remarkable result due to
Seiberg and Witten \cite{Seiberg:1999vs} was that for \emph{gauge
  theories} on noncommutative $\mathbb{R}^4$ there exists a change of
variables such that the $\theta$-expansion of the gauge transformation
in the new variables preserves the order in $\theta$. In other words,
each $\theta$-order of the expanded action is individually
gauge-invariant in the new variables, and one effectively obtains a
commutative gauge theory coupled to a constant external field
$\theta$.

This change of variables  can be traced back
\cite{Bichl:2001yf} to a deeper discussion of Lorentz transformations
\cite{Colladay:1998fq}: In presence of $\theta$ one has to distinguish
between `observer Lorentz transformations', which transform $\theta$
as a Lorentz two-tensor, and `particle Lorentz transformations', which
leave $\theta$ invariant. It turns out that observer Lorentz
transformations are symmetries of the theory whereas particle Lorentz
symmetry is broken. Being (in principle) an observable, the breaking
of particle Lorentz symmetry must be gauge-invariant
\cite{Bichl:2001yf}. This is not automatically the case and demands a
covariant redefinition of the splitting of the observer Lorentz
transformation into particle Lorentz transformation plus
$\theta$-transformation, which is governed by the Seiberg-Witten
differential equations.

It is clear from the construction that the change of variables is
tailored to gauge symmetry. If the action has a second symmetry (apart
from Lorentz symmetries), it is interesting to know whether this
symmetry is $\theta$-diagonalised at the same time with the gauge
symmetry or not. As we show in this letter, supersymmetry (regarded as
a transformation of the components of a noncommutative super vector
field in Wess-Zumino gauge) is \emph{not} diagonalised at the same
time with gauge symmetry. This does not mean that supersymmetry is
lost after the Seiberg-Witten map, it is merely not diagonal in the
$\theta$-order. It would be interesting to search for a change of
variables which $\theta$-diagonalises the supersymmetry
transformations and in turn produces $\theta$-expanded gauge
transformations which mix the $\theta$-orders. This will be done
elsewhere. We stress that the change of variables is unphysical
anyway. The fields become dummy integration variables in the path
integral, and the change of variables merely changes the measure of
integration, which at first order in $\theta$ is a field redefinition
also on quantum level \cite{Grimstrup:2002af}.

The letter is organised as follows. We derive in Sec.~\ref{sec3} the
Seiberg-Witten differential equations (which govern the change of
variables) of super Yang-Mills theory via a covariant splitting of the
observer Lorentz transformations (recalled in Sec.~\ref{sec2}) into
particle Lorentz transformations and a remainder, using the splitting
\cite{Bichl:2001yf} for the gauge field as the starting point. The
Seiberg-Witten differential equations lead to a $\theta$-expansion of
the noncommutative super Yang-Mills action in terms of fields living
on commutative space-time, see Sec.~\ref{sec4}. This $\theta$-expanded
action is automatically invariant under commutative gauge
transformations and commutative Lorentz transformations. It is however
not invariant under commutative supersymmetry transformations.
Instead, the $\theta$-expansion of the noncommutative supersymmetry
transformation yields a symmetry transformation of the
$\theta$-expanded action which extends the usual supersymmetry
transformations by terms of order $n\geq 1$ in $\theta$. This result
implies that the Seiberg-Witten map for super Yang-Mills theory cannot
be expressed in terms of superfields. Some comments on superfields are
given in Sec.~\ref{sec5}.

\section{The noncommutative super Yang-Mills action and its
  symmetries} 
\label{sec2}

The noncommutative $\mathcal{N}{=}1$ super Yang-Mills action is in the
component formulation defined by
\begin{align}
\label{ncaction} 
\Gamma = \int d^4x\;\mathrm{tr}\Big(
-\frac14\hat F^{\mu\nu}\hat F_{\mu\nu} 
+ \mathrm{i} \hat \lambda^a \sigma^\mu_{a\dot{a}} 
\hat{D}_\mu \hat{\bar\lambda}^{\dot{a}}
+ \frac{1}{2} \hat D^2\Big)\;,
\end{align}
where 
\begin{align}
\hat F_{\mu\nu} &:= \partial_\mu \hat A_\nu-\partial_\nu \hat A_\mu 
-\mathrm{i}[\hat A_\mu,\hat A_\nu]_\star\;, 
\\
\hat D_\mu \hat{\bar\lambda}^{\dot{a}} &:= 
\partial_\mu \hat{\bar\lambda}^{\dot a}-
\mathrm{i}[\hat A_\mu,\hat{\bar\lambda}^{\dot a}]_\star\;.
\end{align}
Some useful properties of objects carrying spinor indices
$a,\dot{a}\in\{1,2\}$ are listed in the appendix.  The
$\star$-(anti)commutators of matrix-valued Schwartz class functions
$f,g$ are defined by
\begin{align}
[f,g]_\star &= g\star f - f\star g\;,&
\{f,g\}_\star &= g\star f + f\star g\;,
\end{align}
where the $\star$-product is defined by 
\begin{align}
(f \star g)(x) = \int \!\!d^4 y \int \!\!\frac{d^4k}{(2\pi)^4} 
\,f(x{+}\tfrac{1}{2} \theta{\cdot} k)\, g(x{+}y) \,
\mathrm{e}^{\mathrm{i} k\cdot y}\,,
\end{align}
with $(\theta{\cdot} k)^\mu := \theta^{\mu\nu}k_\nu$, $k\cdot y :=
k_\mu y^\mu$ and $\theta^{\mu\nu}=-\theta^{\nu\mu} \in
M_4(\mathbb{R})$. We consider $\theta^{\mu\nu}$ as the components of a
translation-invariant tensor field. 

The action (\ref{ncaction}) is invariant under gauge transformations
\begin{align}
W^G_{\hat{\omega}}  &=  \int d^4x \;\mathrm{tr} \Big( 
\hat{D}_\mu \hat{\omega} 
\,\frac{\delta}{\delta \hat{A}_\mu} 
-\mathrm{i} [\hat{\bar\lambda}^{\dot a}, \hat\omega]_\star 
\frac{\delta}{\delta \hat{\bar\lambda}^{\dot a}} 
-\mathrm{i}[ \hat\lambda^a,\hat\omega]_\star
\frac{\delta}{\delta \hat\lambda^a}
-\mathrm{i}[\hat D,\hat\omega]_\star\frac\delta
{\delta \hat D}\Big)\;,
\label{WG}
\end{align}
observer Lorentz transformations
\begin{align}
W^T_\tau &= 
\int d^4x\;\mathrm{tr}\Big( 
\partial_\tau \hat{A}_\mu \, \frac{\delta}{\delta \hat{A}_\mu} 
+ \partial_\tau \hat\lambda^a\frac{\delta}{\delta\hat\lambda^a}
+ \partial_\tau \hat{\bar{\lambda}}^{\dot a} \frac{\delta}{
\delta \hat{\bar{\lambda}}^{\dot a}}
+ \partial_\tau \hat{D} \frac{\delta}{\delta \hat{D}}\Big)\;,
\label{WT}
\\
W^R_{\alpha\beta} &:= 
\int d^4x\;\mathrm{tr} \bigg(
\Big(\frac{1}{2} \big\{ x_\alpha, 
\partial_\beta \hat{A}_\mu \big\}_\star 
- \frac{1}{2} \big\{ x_\beta, 
\partial_\alpha \hat{A}_\mu \big\}_\star 
+ g_{\mu\alpha} \hat{A}_\beta
- g_{\mu\beta} \hat{A}_\alpha \Big)
\frac{\delta}{\delta \hat{A}_\mu} 
\nonumber
\\
& \hspace*{5em} + \Big(
\frac12 \{x_\alpha, \partial_\beta \hat\lambda^a\}_\star 
- \frac12 \{x_\beta, \partial_\alpha \hat\lambda^a\}_\star 
+\frac{\mathrm{i}}{2} \hat\lambda^b \sigma_{\alpha\beta\,b}{}^a 
\Big)\frac{\delta}{\delta\hat\lambda^a}
\nonumber
\\
& \hspace*{5em} 
+\Big( \frac12\{x_\alpha,
 \partial_\beta\hat{\bar{\lambda}}^{\dot a}\}_\star
- \frac12\{x_\beta, \partial_\alpha \hat{\bar{\lambda}}^{\dot a}\}_\star
-\frac{\mathrm{i}}{2} 
\bar{\sigma}_{\alpha\beta\,\dot{b}}^{\dot{a}} \hat{\bar{\lambda}}^{\dot b}
\Big)
\frac{\delta}{\delta \hat{\bar{\lambda}}^{\dot a}} 
\nonumber
\\
& \hspace*{5em} 
+ \Big(\frac12\{x_\alpha,\partial_\beta \hat D\}_\star 
- \frac12\{x_\beta,\partial_\alpha \hat D\}_\star \Big)
\frac{\delta}{\delta \hat D}\bigg)
\nonumber
\\
& + \Big(\delta^\mu_\alpha \theta_\beta^{~\nu} 
- \delta^\mu_\beta \theta_\alpha^{~\nu} 
+ \delta^\nu_\alpha \theta^\mu_{~\beta} 
- \delta^\nu_\beta \theta^\mu_{~\alpha}\Big) 
\frac{\partial}{\partial\theta^{\mu\nu}}\;,
\label{WR}
\\
W^D &:= 
\int d^4x\;\mathrm{tr}\bigg( 
\Big(\frac{1}{2} \big\{ x^\delta, 
\partial_\delta \hat{A}_\mu \big\}_\star 
+ \hat{A}_\mu \Big) \,\frac{\delta}{\delta \hat{A}_\mu} 
+ \Big(2\hat D +\frac12\{x^\delta, \partial_\delta \hat D\}_\star\Big)\;
\frac{\delta}{\delta \hat D}
\nonumber
\\
&  \hspace*{5em}
+ \Big(
\frac{3}{2} \hat\lambda^a + \frac{1}{2} \big\{x^\delta, 
\partial_\delta \hat\lambda^a\big\}_\star   \Big) 
\frac{\delta}{\delta \hat\lambda^a} 
+
\Big(\frac{3}{2} \hat{\bar{\lambda}}^{\dot a} 
+ \frac{1}{2} \big\{x^\delta, 
\partial_\delta \hat{\bar{\lambda}}^{\dot a} \big\}_\star \Big)
\frac{\delta}{\delta \hat{\bar{\lambda}}^{\dot a}} 
\bigg)
\nonumber
\\
& - 2 \theta^{\mu\nu} \frac{\partial}{\partial \theta^{\mu\nu}} \;,
\label{WD}
\end{align}
and supersymmetry transformations \cite{Piguet:1996ys}
\begin{align}
W^S_{a} &= \int d^4x\;\mathrm{tr}\Big(
\sigma_{\mu a\dot{a}} \hat{\bar{\lambda}}^{\dot{a}}  
\frac{\delta}{\delta \hat A_\mu}
+ \big(\delta^b_a \hat{D} 
+ \frac{1}{2} \sigma^{\mu\nu\,b}_a \hat{F}_{\mu\nu}\big)
\frac{\delta}{\delta \hat{\lambda}^b}
- \mathrm{i} \sigma^\mu_{a\dot{a}} \hat{D}_\mu \hat{\bar{\lambda}}^{\dot{a}} 
\frac{\delta}{\delta \hat{D}}\Big) \;,
\\
W^{\bar{S}}_{\dot{a}} &= \int d^4x\;\mathrm{tr}\Big(
\hat{\lambda}^a \sigma_{\mu a\dot{a}} \frac{\delta}{\delta \hat A_\mu}
+ \big(\delta^{\dot{b}}_{\dot{a}} \hat{D} 
- \frac{1}{2} \bar{\sigma}^{\mu\nu\,\dot{b}}{}_{\dot{a}} 
\hat{F}_{\mu\nu}\big)
\frac{\delta}{\delta \hat{\bar{\lambda}}^{\dot{b}}}
- \mathrm{i} \hat{D}_\mu \hat{\lambda}^a \sigma^\mu_{a\dot{a}} 
\frac{\delta}{\delta \hat{D}}\Big) \;.
\label{Wsusy}
\end{align}
The partial derivative with respect to $\theta^{\mu\nu}$ has the
property
\begin{align}
\frac{\partial (\hat U \star \hat V)}{\partial \theta^{\mu\nu}} 
&= \frac{\partial \hat U }{\partial \theta^{\mu\nu}} \star \hat V 
+ \hat U \star \frac{\partial \hat V}{\partial \theta^{\mu\nu}} 
+ \frac{\mathrm{i}}{2} (\partial_\mu \hat U) \star 
(\partial_\nu \hat V)\;,
\label{true} 
\end{align}
where the fields
$\hat{A}_\mu,\hat{\lambda}^a,\hat{\bar{\lambda}}^{\dot{a}},
\hat{D}$ must be assumed to be independent of $\theta$.

\section{Seiberg-Witten differential equations}
\label{sec3}

As in (non-supersymmetric) noncommutative Yang-Mills theory
\cite{Bichl:2001yf} we derive the Seiberg-Witten differential
equations via a splitting of the observer Lorentz transformation
$W^R_{\alpha\beta}$ into the covariant particle Lorentz transformation
$\tilde{W}^R_{\hat{\phi};\alpha\beta}$ and a remaining piece
$\tilde{W}^R_{\theta;\alpha\beta}$ involving the Seiberg-Witten differential
equation:
\begin{align}
W^R_{\alpha\beta} &\equiv 
\tilde{W}^R_{\hat{\phi};\alpha\beta} + \tilde{W}^R_{\theta;\alpha\beta} \;,
\label{split}
\\
\tilde{W}^R_{\hat{\phi};\alpha\beta}(\theta^{\mu\nu}) &=0 \;, 
\label{particle}
\\{}
[\tilde{W}^R_{\hat{\phi};\alpha\beta},W^G_{\hat{\omega}}] &=
W^G_{\hat{\omega}'_{\alpha\beta}}\;, & 
[\tilde{W}^R_{\theta;\alpha\beta},W^G_{\hat{\omega}}] &=
W^G_{\hat{\omega}''_{\alpha\beta}}\;.
\label{covariance}
\end{align} 
The motivation for this ansatz is the following. The commutator of
an \emph{observer Lorentz rotation} (\ref{WR}) with a gauge transformation
(\ref{WG}) is again a gauge transformation, 
\begin{align}
[W^R_{\alpha\beta},W^G_{\hat{\omega}}] &=
W^G_{\hat{\omega}_{\alpha\beta}}\;,
\end{align}
for some infinitesimal gauge parameter
$\hat{\omega}_{\alpha\beta}[\hat{\omega}]$. A \emph{particle Lorentz
transformation} is defined as the part of an observer Lorentz
transformation which does not transform the field $\theta^{\mu\nu}$,
see (\ref{particle}). However, one should require that a particle
Lorentz transformation transforms a gauge-invariant quantity into
another gauge-invariant quantity, otherwise the particle Lorentz
transformation cannot be considered as well-defined
\cite{Bichl:2001yf}. It is sufficient to demand (\ref{covariance}) in
order to achieve this property.

To find the sought for splitting we first apply the ansatz of 
ref.~\citelow{Bichl:2001yf} for the Yang-Mills field $\hat{A}_\mu$:
\begin{align}
\tilde{W}^R_{\hat{\phi};\alpha\beta} \hat{A}_\mu &= 
\hat{D}_\mu \hat{\chi}_{\alpha\beta}  
+ \Big(\frac{1}{2} \{\hat{X}_\alpha,\hat{F}_{\beta\mu}\}_\star 
-\frac{1}{2} \{\hat{X}_\beta,\hat{F}_{\alpha\mu}\}_\star 
- W^R_{\alpha\beta} (\theta^{\rho\sigma})
\hat{\Omega}_{\rho\sigma\mu} \Big)\;,
\label{TRA}
\end{align}
where $\hat{X}^\mu=x^\mu+\theta^{\mu\nu}\hat{A}_\nu$ are the covariant
coordinates \cite{Madore:2000en} and $\hat{\Omega}_{\rho\sigma\mu}$ is
a polynomial in covariant quantities such as
$\theta^{\alpha\beta},\hat{F}_{\kappa\lambda},
\hat{D}_{\mu_1}\dots\hat{D}_{\mu_n} \hat{F}_{\kappa\lambda}$,
antisymmetric in $\rho,\sigma$, of power-counting dimension $3$, and
expresses the freedom in the splitting. In the following we set
$\hat{\Omega}_{\rho\sigma\mu}=0$. The parameter
$\hat{\chi}_{\alpha\beta}$ is unchanged and given by
\cite{Bichl:2001yf}
\begin{align}
\hat{\chi}_{\alpha\beta} &= 
\frac{1}{4}\{2 x_\alpha+\theta_\alpha^{~\rho}\hat{A}_\rho,
\hat{A}_\beta\}_\star
-\frac{1}{4}\{2x_\beta+\theta_\beta^{~\rho}\hat{A}_\rho,
\hat{A}_\alpha\}_\star\;.
\label{omega}
\end{align}
Comparing (\ref{TRA}) with the $\hat{A}_\mu$-part of (\ref{WR}) and 
extending this covariantisation to the remaining fields
$\hat{\lambda}^a, \hat{\bar{\lambda}}^{\dot{a}},\hat{D}$ we obtain
from (\ref{WR}) 
\begin{align}
\tilde{W}^R_{\hat{\phi};\alpha\beta} =
W^G_{\hat{\chi}_{\alpha\beta}} + \int d^4x\;\mathrm{tr} &\bigg(
\Big(\frac{1}{2} \big\{ \hat{X}_\alpha, 
\hat{F}_{\beta\mu} \big\}_\star 
- \frac{1}{2} \big\{ \hat{X}_\beta, 
\hat{F}_{\alpha\mu} \big\}_\star \Big)
\frac{\delta}{\delta \hat{A}_\mu} 
\nonumber
\\
& + \Big(
\frac12 \{ \hat{X}_\alpha, \hat{D}_\beta \hat\lambda^a\}_\star 
- \frac12 \{\hat{X}_\beta, \hat{D}_\alpha \hat\lambda^a\}_\star 
+ \frac{\mathrm{i}}{2} \hat\lambda^b \sigma_{\alpha\beta\,b}{}^a \Big)
\frac{\delta}{\delta\hat{\lambda}^a}
\nonumber
\\
& 
+\Big( \frac12\{ \hat{X}_\alpha,
 \hat{D}_\beta\hat{\bar{\lambda}}^{\dot a}\}_\star
- \frac12\{ \hat{X}_\beta, \hat{D}_\alpha 
\hat{\bar{\lambda}}^{\dot a}\}_\star
-\frac{\mathrm{i}}{2} \bar{\sigma}^{\dot{a}}_{\alpha\beta\,\dot{b}} 
\hat{\bar{\lambda}}^{\dot b}\Big)
\frac{\delta}{\delta \hat{\bar{\lambda}}^{\dot a}} 
\nonumber
\\
& 
+ \Big(\frac12\{ \hat{X}_\alpha,\hat{D}_\beta \hat D\}_\star 
- \frac12\{ \hat{X}_\beta,\hat{D}_\alpha \hat D\}_\star \Big)
\frac{\delta}{\delta \hat D}\;,
\label{WTR}
\end{align}
Now it is straightforward to evaluate
\begin{align}
\tilde{W}^R_{\theta;\alpha\beta}=W^R_{\alpha\beta} 
-\tilde{W}^R_{\hat{\phi};\alpha\beta} =
W^R_{\alpha\beta}(\theta^{\rho\sigma})
\frac{d}{d \theta^{\rho\sigma}}\;,
\end{align}
with
\begin{align}
\frac{d}{d \theta^{\rho\sigma}} 
= \frac{\partial}{\partial \theta^{\rho\sigma}} 
+ \int d^4x\;\mathrm{tr}\bigg(\frac{d \hat{A}_\mu}{
d \theta^{\rho\sigma}} 
\frac{\delta}{\delta \hat{A}_\mu}
+   
\frac{d \hat{\lambda}^a}{d \theta^{\rho\sigma}}
\frac{\delta}{\delta \hat{\lambda}^a} 
+ 
\frac{d \hat{\bar{\lambda}}^{\dot a}}{d \theta^{\rho\sigma}} 
\frac{\delta}{\delta \hat{\bar{\lambda}}^{\dot a}} 
+ 
\frac{d \hat{D}}{d \theta^{\rho\sigma}} 
\frac{\delta}{\delta \hat{D}}\bigg)\;,
\end{align}
which yields the Seiberg-Witten differential equations
\begin{align}
\label{SWA}
\frac{d \hat{A}_\mu}{d \theta^{\rho\sigma}}& =  
-\frac{1}{8} \big\{ \hat{A}_\rho, \partial_\sigma \hat{A}_\mu +
\hat{F}_{\sigma\mu} \big\}_\star 
+\frac{1}{8} \big\{ \hat{A}_\sigma, \partial_\rho \hat{A}_\mu +
\hat{F}_{\rho\mu} \big\}_\star~,
\\[1ex]
\frac{d \hat\lambda^a}{d \theta^{\rho\sigma}}& =
-\frac{1}{8} \big\{ \hat{A}_\rho, \partial_\sigma \hat\lambda^a +
\hat D_\sigma \hat\lambda^a\big\}_\star
+\frac{1}{8} \big\{ \hat{A}_\sigma, \partial_\rho \hat\lambda^a 
+\hat D_\rho \hat\lambda^a\big\}_\star~,
\\
\frac{d \hat{\bar\lambda}^{\dot a}}{d \theta^{\rho\sigma}}& =
-\frac{1}{8} \big\{ \hat{A}_\rho, \partial_\sigma \hat{\bar\lambda}^{\dot a}
+\hat D_\sigma \hat{\bar\lambda}^{\dot a}\big\}_\star
+\frac{1}{8} \big\{ \hat{A}_\sigma, \partial_\rho \hat{\bar\lambda}^{\dot a}
+\hat D_\rho \hat{\bar\lambda}^{\dot a}\big\}_\star~,
\\
\frac{d \hat{D}}{d \theta^{\rho\sigma}}& =
-\frac{1}{8} \big\{ \hat{A}_\rho, \partial_\sigma \hat D +\hat D_\sigma 
\hat D\big\}_\star
+\frac{1}{8} \big\{ \hat{A}_\sigma, \partial_\rho \hat D +\hat D_\rho 
\hat D\big\}_\star~.
\label{SWD}
\end{align}
The differential equation (\ref{SWA}) was first found in ref.\ 
\citelow{Seiberg:1999vs}. 

\section{$\theta$-expansion of the action}
\label{sec4}

The differential equations (\ref{SWA})--(\ref{SWD}) are now taken as
the starting point for a $\theta$-expansion of the action,
\begin{align}
\Gamma^{(n)} := \sum_{j=0}^n \frac{1}{j!} 
\theta^{\rho_1\sigma_1} \cdots \theta^{\rho_j\sigma_j} 
\Big(\frac{d^j \Gamma}{d \theta^{\rho_1\sigma_1} \dots 
d \theta^{\rho_j\sigma_j}}\Big)_{\theta=0}\;.
\label{Thexp}
\end{align}
It follows from the the second identity in (\ref{covariance}) that the
$\theta$-expansion (\ref{Thexp}) of the action (\ref{ncaction}) is
invariant under commutative gauge transformations. Using
(\ref{WT})--(\ref{WD}) one also checks the
identity
\begin{align}
\Big[W^{\{T,R,D\}},\theta^{\rho\sigma}
\frac{d}{d \theta^{\rho\sigma}}\Big]=0  
\label{WTRD}
\end{align}
for super Yang-Mills theory, which means that the
$\theta$-expansion of the fields leads to a
commutative action invariant under commutative rotations and
translations and with commutative dilatational symmetry. The identity
(\ref{WTRD}) is a consequence of the fact that $\theta^{\rho\sigma}
\frac{d}{d \theta^{\rho\sigma}}$ is a Lorentz scalar with respect to
observer Lorentz transformations. 

The $\theta$-expansion of (\ref{ncaction}) yields an action which is
\emph{not invariant} under commutative supersymmetry transformations.
Indeed, the commutator of a supersymmetry transformation (\ref{Wsusy})
and a $\theta$-differentiation is given by\footnote{There is of course
  a freedom in the differential equations (\ref{SWA})--(\ref{SWD})
  given by the $\Omega$-terms in (\ref{TRA}) and similarly for the
  other fields. This freedom is not sufficient to obtain a vanishing
  right hand side of (\ref{TS}).}
\begin{align}
\Big[\frac{d}{d \theta^{\rho\sigma}}, W^S_a \Big] &= 
\tilde{W}^G_{\frac{1}{8} \sigma_{\rho\,a\dot{a}} 
\{\hat{A}_\sigma,\smallhatbarlambda{}^{\dot{a}}\}_\star
-\frac{1}{8} \sigma_{\sigma\,a\dot{a}}
\{\hat{A}_\rho,\smallhatbarlambda{}^{\dot{a}}\}_\star}
\nonumber
\\
& + \int d^4x\;\mathrm{tr}\Big( 
\Big(\frac{1}{4} \sigma_{\rho\,a\dot{a}} 
\{\hat{F}_{\sigma\mu},\hat{\bar{\lambda}}^{\dot{a}}\}_\star
-\frac{1}{4} \sigma_{\sigma\,a\dot{a}} 
\{\hat{F}_{\rho\mu},\hat{\bar{\lambda}}^{\dot{a}}\}_\star
\Big) \frac{\delta}{\delta \hat{A}_\mu}
\nonumber
\\
& \qquad\qquad 
+ \Big(\frac{1}{4} \sigma_a^{\mu\nu\,b}
\{\hat{F}_{\mu\rho},\hat{F}_{\nu\sigma}\}_\star 
+\frac{1}{4} \sigma_{\rho\,a\dot{a}} 
[\hat{\bar{\lambda}}^{\dot{a}},\hat{D}_\sigma
\hat{\lambda}^b]_\star 
- \frac{1}{4} \sigma_{\sigma\,a\dot{a}} 
[\hat{\bar{\lambda}}^{\dot{a}},\hat{D}_\rho
\hat{\lambda}^b]_\star \Big) \frac{\delta}{\delta \hat{\lambda}^b}
\nonumber
\\
& \qquad\qquad
+ \Big(\frac{1}{4} \sigma_{\rho\,a\dot{a}} 
[\hat{\bar{\lambda}}^{\dot{a}},\hat{D}_\sigma
\hat{\bar{\lambda}}^{\dot{b}}]_\star 
- \frac{1}{4} \sigma_{\sigma\,a\dot{a}} 
[\hat{\bar{\lambda}}^{\dot{a}},\hat{D}_\rho
\hat{\bar{\lambda}}^{\dot{b}}]_\star \Big) 
\frac{\delta}{\delta \hat{\bar{\lambda}}^{\dot{b}}}
\nonumber
\\
& \qquad\qquad
+ \Big(
\frac{\mathrm{i}}{4} \sigma^\mu_{a\dot{a}} 
 \{\hat{F}_{\sigma\mu}, \hat{D}_\rho
\hat{\bar{\lambda}}^{\dot{a}} \}_\star 
-\frac{\mathrm{i}}{4} \sigma^\mu_{a\dot{a}} 
 \{\hat{F}_{\rho\mu}, \hat{D}_\sigma
\hat{\bar{\lambda}}^{\dot{a}} \}_\star 
\nonumber
\\
& \hspace*{8em} 
+\frac{1}{4} \sigma_{\rho\,a\dot{a}} 
\{\hat{\bar{\lambda}}^{\dot{a}}, \hat{D}_\sigma \hat{D} \}_\star
-\frac{1}{4} \sigma_{\sigma\,a\dot{a}} 
\{\hat{\bar{\lambda}}^{\dot{a}}, \hat{D}_\rho \hat{D} \}_\star
\Big) \frac{\delta}{\delta \hat{D}}\;,
\label{TS}
\end{align}
where the gauge transformation with respect to a fermionic parameter
$\tilde{\omega}$ is defined by
\begin{align}
\tilde{W}^G_{\tilde{\omega}} &=  \int d^4x \;\mathrm{tr} \Big( 
\hat{D}_\mu \hat{\omega} 
\,\frac{\delta}{\delta \hat{A}_\mu} 
+\mathrm{i} \{\hat{\bar\lambda}^{\dot a}, \tilde\omega\}_\star 
\frac{\delta}{\delta \hat{\bar\lambda}^{\dot a}} 
+\mathrm{i}\{ \hat\lambda^a,\tilde\omega\}_\star
\frac{\delta}{\delta \hat\lambda^a}
-\mathrm{i}[\hat D,\tilde\omega]_\star\frac\delta
{\delta \hat D}\Big)\;.
\label{WGF}
\end{align}
The action (\ref{ncaction}) is invariant under the transformation
(\ref{WGF}). It follows now from (\ref{Thexp}) that the
$\theta$-expansion of (\ref{ncaction}) is invariant under
the transformation
\begin{align}
W^{S,comm}_a &= \big(W^S_a\big)_{\theta=0}
+ \sum_{n=1}^\infty \frac{1}{n!} \theta^{\rho_1\sigma_1} \cdots 
\theta^{\rho_n\sigma_n} \Big(
\Big[\frac{d}{d\theta^{\rho_1\sigma_1}},\Big[ \dots \Big[
\frac{d}{d\theta^{\rho_n\sigma_n}},W^S_a\Big]\dots\Big]\Big]
\Big)_{\theta=0}\;,
\label{WSc}
\end{align}
which due to $[\frac{d}{d\theta},W^S_a]\neq 0$ is different from the
commutative supersymmetry transformation $\big(W^S_a\big)_{\theta=0}$.
In other words, the Seiberg-Witten map does not diagonalise the
$\theta$-expansion of the noncommutative supersymmetry transformation,
see also the remarks in the Introduction. The first terms of
(\ref{WSc}) read
\begin{align}
W^{S,comm}_{a} &= \int d^4x\;\mathrm{tr}\Big(
\Big(\sigma_{\mu a\dot{a}} \bar{\lambda}^{\dot{a}}  
+ \frac{1}{2} \theta^{\rho\sigma} \sigma_{\rho\,a\dot{a}} 
\{F_{\sigma\mu},\bar{\lambda}^{\dot{a}}\}
\Big) \frac{\delta}{\delta A_\mu}
+ \Big(\frac{1}{2} \theta^{\rho\sigma} \sigma_{\rho\,a\dot{a}} 
[\bar{\lambda}^{\dot{a}},D_\sigma \bar{\lambda}^{\dot{b}}]\Big) 
\frac{\delta}{\delta \bar{\lambda}^b}
\nonumber
\\
& \qquad 
+ \Big(\delta^b_a D 
+ \frac{1}{2} \sigma^{\mu\nu\,b}_a F_{\mu\nu}
+\frac{1}{4} \theta^{\rho\sigma} \sigma_a^{\mu\nu\,b}
\{F_{\mu\rho},F_{\nu\sigma}\}
+ \frac{1}{2} \theta^{\rho\sigma} \sigma_{\rho\,a\dot{a}} 
\{F_{\sigma\mu},\bar{\lambda}^{\dot{a}}\}
\Big) \frac{\delta}{\delta \lambda^b}
\nonumber
\\
& \qquad 
+ \Big(- \mathrm{i} \sigma^\mu_{a\dot{a}} D_\mu \bar{\lambda}^{\dot{a}} 
+ \frac{\mathrm{i}}{2} \theta^{\rho\sigma} \sigma^\mu_{a\dot{a}} 
 \{F_{\sigma\mu}, D_\rho \bar{\lambda}^{\dot{a}} \}
+\frac{1}{2} \theta^{\rho\sigma} \sigma_{\rho\,a\dot{a}} 
\{\bar{\lambda}^{\dot{a}}, D_\sigma D \}
\Big) \frac{\delta}{\delta D} \Big) + \mathcal{O}(\theta^2)\;.
\end{align}
Similar formulae exist for the anti-supersymmetry transformation
$W^{\bar{S}}_{\dot{a}}$. An analogous result for $U(1)$-theory with 
general $\theta^{\mu\nu}$ has also been obtained\cite{Paban}. 

At order $n=0$ in $\theta$ the expansion of (\ref{ncaction}) is obviously
the standard super Yang-Mills action
\begin{align} 
\Gamma^{(0)} 
= \int d^4x\;\mathrm{tr}\Big(-\frac14 F^{\mu\nu} F_{\mu\nu} 
+ \mathrm{i} \lambda^a \sigma^\mu_{a\dot{a}} D_\mu \bar\lambda^{\dot{a}} 
+ \frac 12 D^2\Big)\;,
\end{align} 
where $\phi=\hat{\phi}|_{\theta=0}$ for $\phi \in
\{A_\mu,\lambda^a,\bar{\lambda}^{\dot{a}},D\}$. At first order in
$\theta$ one finds 
\begin{align}
\Gamma^{(1)} = \Gamma^{(0)} - \frac{1}{2}  
\int d^4x\;\mathrm{tr}
\Big( & 
\theta^{\rho\sigma} F_{\rho\sigma}
\Big( -\frac{1}{4} F_{\mu\nu}F^{\mu\nu} 
+\frac{\mathrm{i}}{2} \bar{\lambda}^{\dot{a}}
\bar{\sigma}^\mu_{\dot{a}a} D_\mu \lambda^a 
+ \frac{\mathrm{i}}{2} \lambda^a 
\sigma^\mu_{a\dot{a}} D_\mu \bar{\lambda}^{\dot{a}}
+\frac{1}{2} D^2\Big)
\nonumber
\\
& + \theta^{\rho\sigma} F_{\mu\rho}F_{\nu\sigma}F^{\mu\nu}
+ \theta^{\rho\sigma}  F_{\mu\rho} \big( 
\mathrm{i}\bar{\lambda}^{\dot{a}}
\bar{\sigma}^\mu_{\dot{a}a} D_\sigma \lambda^a 
+ \mathrm{i}\lambda^a 
\sigma^\mu_{a\dot{a}} D_\sigma \bar{\lambda}^{\dot{a}}\big)
\Big)\;.
\label{G1}
\end{align} 
The $\theta$-expanded action (\ref{G1}) could be further analysed, for
instance with respect to new decay channels of supersymmetric
particles---in a similar manner as investigations of models without
supersymmetry \cite{Behr:2002wx}.

\section{Remarks on the superspace formalism}
\label{sec5}

The most compact way to formulate supersymmetric theories is to use
the superfield formalism. The above considered fields
$\hat{A}_\mu,\hat{\lambda}^a,\hat{\bar{\lambda}}^{\dot{a}},\hat{D}$ of
super Yang-Mills theory can be regarded as components of the
superfield
\begin{align}
\hat{\phi} &= \hat{C} + \hat{\chi}^a \theta_a 
+ \bar{\theta}_{\dot{a}} \hat{\bar{\chi}}^{\dot{a}}
+ \theta^a \theta_a \hat M 
+ \bar\theta_{\dot{a}} \bar{\theta}^{\dot{a}} \hat{\bar{M}}
\nonumber
\\
& -2 \theta^a \sigma^\mu_{a \dot{a}} \bar\theta^{\dot{a}} \hat{A}_\mu  
- 2\bar\theta_{\dot a} \hat{\bar\lambda}^{\dot a} \theta^a \theta_a 
- 2\hat\lambda^a \theta_a \bar\theta_{\dot{a}} \bar\theta^{\dot{a}} 
- \theta^a \theta_a \bar\theta_{\dot{a}} \bar\theta^{\dot{a}} 
\hat{D}\;.
\end{align}
The anticommuting variables $\theta^a,\bar{\theta}^{\dot{a}}$ should
not be confused with the noncommutativity parameter $\theta^{\mu\nu}$.
The Wess-Zumino gauge consists in setting the components $\hat{C},
\hat{\chi}^a, \hat{\bar{\chi}}^{\dot{a}},\hat{M},\hat{\bar M}$ equal
to zero. One has $\hat{\phi} \star \hat{\phi} \star \hat{\phi}=0$ in
this gauge. For details about the superfield formalism we refer to
ref.~\citelow{Piguet:ug}.

Due to $[\frac{d}{d \theta},W^S_a]\neq 0$, see (\ref{TS}), a
Seiberg-Witten map in terms of superfields cannot exit. All one can do
is to write the previous formulae in a more compact form, in which the
super vector field is understood to be in Wess-Zumino gauge.  The
gauge transformations and observer Lorentz transformations can be
written in the compact form
\begin{align}
W^G_{\hat\omega} &=\int d^4x\;
\Big(-2\theta^a\sigma^\mu_{a\dot{a}}\bar\theta^{\dot{a}} 
\partial_\mu\hat\omega
 -\mathrm{i}[\hat\phi,\hat\omega]_\star)\Big)
\frac\delta{\delta\hat\phi}\;,
\label{SWG}
\\
W^T_{\tau}  &:= \int d^4x\;\mathrm{tr} 
\Big(\partial_\tau \hat\phi \;\frac{\delta}{\delta \hat\phi}\Big) ,
\label{TPHI}
\\
W^R_{\alpha\beta} &:= 
\int d^4x\;\mathrm{tr}\Big( 
\Big(\frac{1}{2} \big\{ x_\alpha, 
\partial_\beta \hat{\phi} \big\}_\star 
- \frac{1}{2} \big\{ x_\beta, 
\partial_\alpha \hat{\phi} \big\}_\star 
+ \Sigma_{\alpha\beta}\hat\phi\Big)
\frac{\delta}{\delta \hat{\phi}} \Big) 
\nonumber
\\
& + \Big(\delta^\mu_\alpha \theta_\beta^{~\nu} 
- \delta^\mu_\beta \theta_\alpha^{~\nu} 
+ \delta^\nu_\alpha \theta^\mu_{~\beta} 
- \delta^\nu_\beta \theta^\mu_{~\alpha}\Big) 
\frac{\partial}{\partial\theta^{\mu\nu}}  \;,
\label{RPHI}
\\
W^D &= \int d^4x\;\mathrm{tr}\Big( 
\frac{1}{2} \big\{ x^\delta, \partial_\delta \hat{\phi} \big\}_\star 
\frac{\delta}{\delta \hat{\phi}}\Big) - 2 \theta^{\mu\nu}
\frac{\partial}{\partial \theta^{\mu\nu}}\;.
\end{align}
Here $\Sigma_{\alpha\beta}=-\frac{\mathrm{i}}{2} \theta^a 
\sigma_{\alpha\beta\,a}{}^b
\frac{\partial}{\partial \theta^b}+\frac{\mathrm{i}}{2} \bar{\theta}_{\dot{a}}
\bar{\sigma}_{\alpha\beta\,\dot{b}}^{\dot{a}} \frac{\partial}{\partial
  \bar{\theta}_{\dot{b}}}$ is the spin operator for the superfield. The
covariant particle Lorentz rotation reads
\begin{align}
\tilde W^R_{\hat{\phi};\alpha\beta} &:= W^G_{\hat\chi_{\alpha\beta}}
+ \int d^4x\;\mathrm{tr}\Big( 
\Big(\frac{1}{2} \big\{ \hat X_\alpha, 
\hat F_\beta \big\}_\star
- \frac{1}{2} \big\{ \hat X_\beta, 
\hat F_\alpha\big\}_\star 
+ \Sigma_{\alpha\beta}\big( \hat\phi
+2  \theta^a\sigma^\mu_{a\dot{a}}\bar\theta^{\dot{a}} 
\hat A_\mu\big)\Big)
\frac{\delta}{\delta \hat{\phi}} \Big) \;,
\label{RCPHI}
\end{align}
where $\hat\chi_{\alpha\beta}$ is given by (\ref{omega}) and 
\begin{align}
\hat F_\sigma &:= \partial_\sigma\hat \phi +
2 \theta^a\sigma^\mu_{a\dot{a}}\bar\theta^{\dot{a}} 
\partial_\mu\hat A_\sigma 
- \mathrm{i} [\hat A_\sigma,\hat\phi]_\star\;.
\end{align}
This object, resembling the usual field strength tensor $F_{\mu\nu}$,
transforms covariantly under supergauge transformations (\ref{SWG}).
The calculation of the Seiberg-Witten expansion is straightforward and
yields
\begin{align}
\frac{d \hat \phi}{d \theta^{\rho\sigma}}& =
-\frac{1}{8} \big\{ \hat{A}_\rho, \partial_\sigma\hat\phi + 
\hat F_\sigma\big\}_\star
+\frac{1}{8} \big\{ \hat{A}_\sigma, \partial_\rho \hat \phi +
\hat F_\rho \big\}_\star\;.
\end{align}

\section{Conclusion} \label{con}

Following previous ideas \cite{Bichl:2001yf,Grimstrup:2001ja} we have
derived the Seiberg-Witten map for noncommutative super Yang-Mills
theory in Wess-Zumino gauge via the splitting of the observer Lorentz
transformation into a covariant particle Lorentz transformation and a
remainder, which directly leads to the Seiberg-Witten differential
equations. We have also computed the $\theta$-expansion of the
noncommutative super Yang-Mills action, up to first order in $\theta$.
Each $\theta$-order of the action is individually invariant under
commutative gauge transformations. In contrast, the $\theta$-expansion
of the supersymmetry transformation \emph{differs} from the
commutative supersymmetry transformations by terms of order $n\geq 1$
in $\theta$. For this reason the Seiberg-Witten map cannot be
expressed in terms of superfields.

\begin{appendix}

\section{Useful formulae}

\renewcommand{\theequation}{\Alph{section}.\arabic{equation}}
\setcounter{equation}{0}

Spinor indices $a,\dot{a}\in\{1,2\}$ are shifted by the antisymmetric 
metric
$\varepsilon^{ab}=-\varepsilon^{ba},\varepsilon^{\dot{a}\dot{b}}
=-\varepsilon^{\dot{b}\dot{a}}$ according to
\begin{align}
\chi_a &= \varepsilon_{ab} \chi^b\;, &
\bar\chi^{\dot{a}} &= \varepsilon^{\dot{a}\dot{b}} \bar\chi_{\dot{b}}\;.
\end{align}
Note that spinors are anticommuting, 
\begin{align}
\chi^a \eta_a = - \chi_a \eta^a = \eta^a \chi_a = - \eta_a \chi^a
\;,\qquad  
\bar\chi_{\dot{a}} \bar\eta^{\dot{a}} 
= - \bar\chi^{\dot{a}} \bar\eta_{\dot{a}} 
= \bar\eta_{\dot{a}} \bar\chi^{\dot{a}} 
= -\bar\chi^{\dot{a}} \bar\eta_{\dot{a}} \;.
\end{align}
The $2{\times} 2$ $\sigma$-matrices are given by
\begin{align}
\sigma^\mu_{a\dot{a}} &=\big(1,\vec{\sigma}\big)_{a\dot{a}}\;, &
\bar{\sigma}^{\mu\,\dot{a}a} 
&=\big(1,-\vec{\sigma}\big)^{\dot{a}a}\;, & 
\sigma^\mu_{a\dot{a}}&=\bar{\sigma}^\mu_{\dot{a}a}\;,
\end{align}
where $\vec{\sigma}$ denotes the three Pauli matrices. The
$\sigma$-matrices satisfy 
\begin{align}
\sigma^\mu_{a\dot{a}} \bar{\sigma}^{\nu\,\dot{a}b} 
&= g^{\mu\nu} \delta_a^{~b} - \mathrm{i}\sigma^{\mu\nu\,b}_a\;, 
\\
\bar{\sigma}^{\mu\,\dot{a}a} \sigma^\nu_{a\dot{b}} 
&= g^{\mu\nu} \delta^{\dot{a}}_{~\dot{b}} 
- \mathrm{i}\bar{\sigma}^{\mu\nu\,\dot{a}}{}_{\dot{b}}\;,
\\
\sigma^\mu_{a\dot{a}} \bar{\sigma}^{\nu\,\dot{a}b}
\sigma^\rho_{b\dot{b}} &= 
g^{\mu\nu} \sigma^\rho_{a\dot{b}}
+g^{\nu\rho} \sigma^\mu_{a\dot{b}}
-g^{\rho\mu} \sigma^\nu_{a\dot{b}}
- \mathrm{i} \epsilon^{\mu\nu\rho\lambda} \sigma_{\lambda\,a\dot{b}}
\;, 
\\
\bar{\sigma}^{\mu\,\dot{a}a} \sigma^\nu_{a\dot{b}} 
\bar{\sigma}^{\rho\,\dot{b}b}
&= g^{\mu\nu} \bar{\sigma}^{\rho\,\dot{a}b}
+ g^{\nu\rho} \bar{\sigma}^{\mu\,\dot{a}b}
- g^{\rho\mu} \bar{\sigma}^{\nu\,\dot{a}b}
+ \mathrm{i} \epsilon^{\mu\nu\rho\lambda} 
\bar{\sigma}_\lambda^{\dot{a}b}\;, 
\\
\sigma^\mu_{a\dot{a}}\sigma_{\mu\,b\dot{b}}
& =2\varepsilon_{ab}\varepsilon_{\dot{a}\dot{b}} \;,
\end{align}
with $\sigma^{\mu\nu\,b}_a= -\sigma^{\nu\mu\,b}_a$ and 
$\bar{\sigma}^{\mu\nu\,\dot{a}}{}_{\dot{b}}
=-\bar{\sigma}^{\nu\mu\,\dot{a}}{}_{\dot{b}}$.

\end{appendix}


\begin{thebibliography}{9}

\bibitem{Seiberg:1999vs} N.~Seiberg and E.~Witten, 
  JHEP {\bf 9909} (1999) 032 [hep-th/9908142].

\bibitem{Chepelev:2001hm} I.~Chepelev and R.~Roiban, 
 JHEP {\bf 0103} (2001)
  001 [arXiv:hep-th/0008090].

\bibitem{Bichl:2001cq}
A.~Bichl, J.~Grimstrup, H.~Grosse, L.~Popp, M.~Schweda and R.~Wulkenhaar,
JHEP {\bf 0106} (2001) 013
[arXiv:hep-th/0104097].

\bibitem{Wulkenhaar:2001sq}
R.~Wulkenhaar,
JHEP {\bf 0203} (2002) 024
[arXiv:hep-th/0112248].

\bibitem{Bichl:2001yf} A.~A.~Bichl, J.~M.~Grimstrup, H.~Grosse,
  E.~Kraus, L.~Popp, M.~Schweda and R.~Wulkenhaar, 
 ``Noncommutative
  Lorentz symmetry and the origin of the Seiberg-Witten map,''
  hep-th/0108045.

\bibitem{Colladay:1998fq}
D.~Colladay and V.~A.~Kostelecky,
Phys.\ Rev.\ D {\bf 58} (1998) 116002
[hep-ph/9809521].

\bibitem{Grimstrup:2002af}
J.~M.~Grimstrup and R.~Wulkenhaar,
Eur.\ Phys.\ J.\ C {\bf 26} (2002) 139
[arXiv:hep-th/0205153].

\bibitem{Piguet:1996ys} O.~Piguet, 
 ``Supersymmetry, supercurrent, and scale invariance,'' 
 arXiv:hep-th/9611003.
  
\bibitem{Madore:2000en} J.~Madore, S.~Schraml, P.~Schupp and J.~Wess,
 Eur.\ Phys.\ J.\ C {\bf 16} (2000) 161 [hep-th/0001203].

\bibitem{Paban} S.~Paban, S.~Sethi and M.~Stern,
``Non-commutativity and Supersymmetry,''
arXiv:hep-th/0201259.

\bibitem{Behr:2002wx} W.~Behr, N.~G.~Deshpande, G.~Duplancic,
  P.~Schupp, J.~Trampetic and J.~Wess, 
  ``The $Z \to \gamma \gamma$,
  $gg$ decays in the noncommutative standard model,''
  arXiv:hep-ph/0202121.

\bibitem{Piguet:ug} O.~Piguet and K.~Sibold, 
 ``Renormalized
  Supersymmetry. The Perturbation Theory Of $N=1$ Supersymmetric
  Theories In Flat Space-Time,'' 
 {\it Boston, USA: Birkhaeuser (1986)
    346 P. (Progress In Physics, 12)}.

\bibitem{Grimstrup:2001ja}
J.~M.~Grimstrup, H.~Grosse, E.~Kraus, L.~Popp, M.~Schweda and R.~Wulkenhaar,
``Noncommutative spin-1/2 representations,''
arXiv:hep-th/0110231.

\end{thebibliography}
\end{document}